\documentstyle[12pt,psfig]{article}
\def\be{\begin{equation}}
\def\ee{\end{equation}}
\def\bea{\begin{eqnarray}}
\def\eea{\end{eqnarray}}
\def\lsim{\:\raisebox{-0.5ex}{$\stackrel{\textstyle<}{\sim}$}\:}
 
\begin{document}
\begin{flushright}
TIFR/TH/01-36 \\
CERN/TH/01-246\\
\end{flushright}
\bigskip
\begin{center}
{\Large\bf Using Tau Polarization to Sharpen up the SUGRA Signal at
Tevatron} \\[2cm] 
{\large Monoranjan Guchait$^a$ and D.P. Roy$^{a,b}$} \\[1cm]
$^a$ Theoretical Physics Division, CERN \\
CH-1211, Geneva 23, Switzerland \\
$^b$ Tata Institute of Fundamental Research \\
Homi Bhabha Road, Mumbai 400 005, INDIA
\end{center}

\vspace{1cm}

\begin{center}
{\large\bf Abstract}
\end{center}
\bigskip

The most promising source of SUGRA signal at the Tevatron collider is
the pair-production of electroweak gauginos, followed by their
leptonic decay.  In the parameter range corresponding to dominant
leptonic decay of these gauginos one or more of the leptons are
expected to be $\tau$ with $P_\tau \simeq +1$.  This polarization can
be effectively used to distinguish the signal from the background in
the 1-prong hadronic decay channel of $\tau$ by looking at the
fractional $\tau$-jet momentum carried by the charged prong.

\newpage

The LEP limit of chargino mass $> 90$ GeV corresponds to a gluino mass
$> 300$ GeV in the minimal SUGRA model \cite{one}, which puts them beyond the
discovery reach of the Tevatron collider.  Thus the most promising
source of SUGRA signal at Tevatron seems to be the pair production of
electroweak gauginos,  $\tilde W^+_1 \tilde W^-_1$ and $\tilde W^\pm_1
\tilde Z_2$.  The leptonic decays of $\tilde W_1$ and $\tilde Z_2$
into the LSP $(\tilde Z_1)$ result in clean dilepton and trilepton
final states with a significant missing-$E_T$ $({E\!\!\!/}_T)$ and
very little hadronic jet activity.  Recently there has been a good
deal of interest in these processes as the main signatures of the
SUGRA model at Tevatron \cite{two}-\cite{five}.  The parameter space
of particular interest to this signature is one where the lighter
(right-handed) sleptons lye below the $\tilde W_1$ and $\tilde Z_2$
masses, resulting in very large leptonic branching fractions of these
gauginos.  This corresponds to two regions of the SUGRA parameter
space -- i.e. I) $m_0$ significantly less than $m_{1/2}$ $(m_0 \sim
{1\over2} m_{1/2})$, implying $m_{\tilde\ell_R,\tilde\tau_R} \lsim
m_{\tilde W_1,\tilde Z_2}$ at any value of $\tan\beta$, and II) $m_0
\sim m_{1/2}$ at large $\tan\beta$, implying $m_{\tilde \tau_R} \lsim
m_{\tilde W_1,\tilde Z_2}$, where $\ell$ denotes electron and muon.
In the 1st case one expects a $\ell^+ \ell^- \tau$ signature from
$\tilde Z_2 \tilde W_1$ decay, since $\tilde W_1 \rightarrow \tau
\nu_\tau \tilde Z_1$ via the larger L-R mixing in the $\tilde\tau$
sector due to the larger $\tau$ mass.  In the 2nd case one expects
$\tau\tau$ and $\tau\tau\tau$ 
signatures from $\tilde W_1 \tilde W_1$ and $\tilde Z_2 \tilde W_1$
decays respectively.  The presence of one or more $\tau$ leptons in
the final state means that the $\tau$ channel is
expected to play a very important role in superparticle search at
Tevatron, particularly in the minimal SUGRA model
\cite{two}-\cite{five}. 

The minimal SUGRA model predicts the polarization of $\tau$ resulting
from the above $\tilde\tau$ decay to be $= +1$ to a good
approximation, as we shall see below.  The purpose of this note is to
use this $\tau$ polarization $(P_\tau = +1)$ to sharpen the
distinction between the SUSY signal and the SM background.  It has
been shown in the context of charged Higgs boson search in the $H^\pm
\rightarrow \tau \nu$ channel that in the 1-prong hadronic $\tau$-jet
the $P_\tau = +1$ signal from $H^\pm$ decay can be effectively
distinguished from the $P_\tau = -1$ background from $W^\pm$ via the
sharing of the jet energy between the 
charged pion and the accompanying neutrals \cite{six}-\cite{seven}.
This has been confirmed now by detailed simulation studies for both
Tevatron and LHC.  We shall use a similar strategy here to distinguish
the SUSY signal from the SM background in the 1-prong hadronic
$\tau$-jet channels.  In particular we shall see that the $P_\tau = +1$
signal can be effectively separated from the $P_\tau = -1$ background
as well as the fake $\tau$ background from QCD jets by requiring the
charged track to carry $> 80\%$ of the jet energy-momentum.

We shall concentrate on the 1-prong hadronic decay channel of $\tau$,
which is best suited for $\tau$ identification.  It accounts for 80\%
of hadronic $\tau$ decay and 50\% of its total decay width.  The main
contributors to the 1-prong hadronic decay are 
\be
\tau^\pm \rightarrow \pi^\pm \nu (12.5\%), \ \rho^\pm \nu (26\%), \
a^\pm_1 \nu (7.5\%),
\label{one}
\ee
where the branching fractions for $\pi$ and $\rho$ include the small
$K$ and $K^\star$ contributions respectively \cite{one}, which have
identical polarization effects.  Together they account for 90\% of the
1-prong hadronic decay. The CM angular distribution of $\tau$ decay
into $\pi$ or a vector meson $v$ $(= \rho,a_1)$ is simply given in
terms of its polarization as
\be
{1 \over \Gamma_\pi} {d\Gamma_\pi \over d\cos\theta} = {1\over2} (1 +
P_\tau \cos\theta),
\label{two}
\ee
\be
{1 \over \Gamma_v} {d\Gamma_{v L,T} \over d\cos\theta} =
{{1\over2} m^2_\tau, m^2_v \over m^2_\tau + 2m^2_v} (1 \pm P_\tau
\cos\theta), 
\label{three}
\ee
where $L,T$ denote the longitudinal and transverse polarization states
of the vector meson.  The fraction $x$ of the $\tau$ lab. momentum carried
by its decay meson is
related to the angle $\theta$ via
\be
x = {1\over2} (1 + \cos\theta) + {m^2_{\pi,v} \over 2m^2_\tau} (1 -
\cos\theta), 
\label{four}
\ee
where we have neglected the $\tau$ mass relative to its lab. momentum
(collinear approximation). The only measurable $\tau$ momentum is the
product $xp_\tau = p_{\tau-{\rm jet}}$, i.e. the visible momentum of the
$\tau$-jet.  It is clear from eqs. (\ref{two}) -
(\ref{four}) that the hard part of the $\tau$-jet, which is
responsible for $\tau$ identification, is dominated by
$\pi,\rho_L,a_{1L}$ for the $P_\tau = +1$ signal, while it is
dominated by $\rho_T,a_{1T}$ for the $P_\tau = -1$ background.  The
two can be distinguished by exploiting the fact that the transverse $\rho$ and
$a_1$ decays favour even sharing of momentum among the decay pions,
while the longitudinal $\rho$ and $a_1$ decays favour uneven sharing,
where the charged pion carries either very little or most of the
momentum.  It is easy to derive this quantitatively for $\rho$ decay.
But one has to assume a dynamical model for $a_1$ decay to get a
quantitative result.  We shall assume the model of ref. \cite{eight},
based on conserved axial vector current approximation, which provides
a good description to the $a_1 \rightarrow 3\pi$ data.  A detailed
account of the $\rho$ and $a_1$ decay formalisms including finite
width effects can be found in \cite{six},\cite{nine}.  A simple
FORTRAN code for 1-prong hadronic decay of Polarized $\tau$ based on
these formalisms can be obtained from one of the authors (D.P. Roy).
It gives the distribution of the $\tau$ momentum among the decay pions
in the 1-prong hadronic decay mode into one charged and any numbers of neutral
pions,
\be
\tau^\pm \rightarrow \pi^\pm(\pi^0s)\nu 
\ee
in terms of the $\pi$, $\rho$ and $a_1$ contributions of eq.(\ref{one}).
This is a 2-step process. First it gives the fraction of the $\tau$ 
momentum imparted to the visible $\tau$-jet (i.e. $\pi$, $\rho_L$, $\rho_T$,
$a_{1L}$ or $a_{1T}$) via eqs. (\ref{two}) - (\ref{four}). Then it 
determines how this visible $\tau$-jet momentum is shared between the 
decay pions using the $\rho_{L,T}$ and $a_{1L,T}$ decay formalism of
refs. \cite{six}, \cite{nine}. 

As we shall see below the two polarization states predict distinctive
distributions in $R=p_{\pi^\pm}/p_{\tau-jet}$, i.e. the 
fraction of the visible $\tau$-jet momentum carried by the charged 
prong. This can be obtained by combining the charged prong momentum 
measurement in the tracker with the calorimetric energy deposit
of the $\tau$-jet.

As specific examples of the two regions of interest in the SUGRA
parameter space mentioned above, we have chosen two points
representing the cases I and II, and evaluated the corresponding SUSY
spectra using the ISASUGRA code -- version 7.48 \cite{ten}.  The
resulting $\tilde W_1, \tilde Z_2, \tilde Z_1$ and the slepton masses
are shown in the two rows of Table 1 along with the $\tilde\tau$
mixing angle, where
\be
\tilde\tau_1 = \tilde\tau_R \sin\theta_\tau + \tilde\tau_L
\cos\theta_\tau.
\label{five}
\ee
It may be noted here that the Polarization of the $\tau$ resulting
from the $\tilde\tau_1 \rightarrow \tau \tilde Z_1$ decay is
\be
P_\tau \simeq {4\sin^2 \theta_\tau - \cos^2 \theta_\tau \over 4\sin^2
\theta_\tau + \cos^2 \theta_\tau},
\label{six}
\ee
since $\tilde Z_1 \simeq \tilde B$ in the minimal SUGRA model and
$\tau_R$ has twice as large a hypercharge as $\tau_L$ \cite{eleven}.
For the mixing angles of Table 1, $\cos\theta_\tau = 0.19 (0.53)$, we
get $P_\tau = 0.98 (0.85)$.  Hence the $\tau$ polarization is $\simeq
+1$ to a good approximation over a wide range of the relevant SUGRA
parameters, notably $\tan\beta$.

\begin{enumerate}
\item[{}] Table 1. The light sparticle masses of the minimal SUGRA
model for (I)~$m_{1/2} = 175$, $m_0 = 70$, $\tan\beta = 5$ (top row) and
(II)~$m_{1/2} = 160$, $m_0 = 150$, $\tan\beta = 40$ (bottom row), where all
the masses are in GeV unit.  For both cases $A_0 = 0$ and ${\rm sign}(\mu)
= +{\rm ve}$.
\end{enumerate}
\[
\begin{tabular}{|c|c|c|c|c|c|c|c|c|c|}
\hline
$\tilde W_1$ & $\tilde Z_2$ & $\tilde Z_1$ & $\tilde\nu_\ell$ &
$\tilde\ell_L$ & $\tilde\ell_R$ & $\tilde\nu_\tau$ & $\tilde\tau_2$ &
$\tilde\tau_1$ & $\cos\theta_\tau$ \\
\hline
&&&&&&&&& \\
116 & 120 & 66 & 130 & 151 & 106 & 130 & 153 & 104 & 0.19 \\
&&&&&&&&& \\
110 & 112 & 63 & 178 & 195 & 168 & 164 & 207 & ~97 & 0.53 \\
&&&&&&&&& \\
\hline
\end{tabular}
\]

We have estimated the signal and background cross-sections for the
above two cases using a parton level Monte Carlo programme.  To
simulate detector resolution we have applied Gaussian smearing on the
jet and lepton $p_T$ with $(\sigma(p_T)/p_T)^2 = (0.6/\sqrt{p_T})^2 +
(0.04)^2$ and $(0.15/\sqrt{p_T})^2 + (0.01)^2$ respectively.  The
${E\!\!\!/}_T$ is evaluated from the vector sum of the lepton and jet
$p_T$ after resolution smearing.  The main results for the two cases
are presented below.

\subsection*{I.~$m_0 \sim {1\over2} m_{1/2}$ ($\ell^+\ell^-
\tau$ signal):}  
The sparticle spectrum of the top row of Table 1 imply that the
dominant decay modes of $\tilde Z_2$ and $\tilde W_1$ are
\be
\tilde Z_2 \rightarrow \ell \tilde \ell_R \rightarrow \ell^+ \ell^-
\tilde Z_1,
\label{seven}
\ee
\be
\tilde W_1 \rightarrow \nu_\tau \tilde\tau_1 \rightarrow \tau \nu_\tau
\tilde Z_1,
\label{eight}
\ee
with branching fractions $\simeq 2/3$ and $1$ respectively.  Thus one
expects a distinctive $\ell^+ \ell^- \tau$ signal accompanied by a
significant ${E\!\!\!/}_T$ from $\tilde W_1 \tilde Z_2$ production.
Moreover this signal is expected to hold over a wide range of
$\tan\beta$, since the production cross-section as well as the above decay
branching fractions are insensitive to this parameter.  Note that the
right-handed slepton masses of Table 1 are fairly close to the $\tilde
W_1, \tilde Z_2$ masses due to the LEP limit on $m_{\tilde\ell,\tilde\tau}$
\cite{one}.  Hence the lepton from the $\tilde Z_2 \rightarrow \ell
\tilde\ell_R$ decay is expected to be relatively soft.  We have
therefore imposed a modest but realistic $p_T$ cut on the softer
lepton.  The cuts are
\[
p^{\ell_1}_T > 15 \ {\rm GeV}, \ p^{\ell_2}_T > 10 \ {\rm GeV},
p_T^{\tau-{\rm jet}} > 15 \ {\rm GeV}, {E\!\!\!/}_T > 20 \ {\rm GeV},
\]
\be
|\eta_{\ell_1,\ell_2,\tau-{\rm jet}}| < 2.5, \ \phi_{\ell_1 \ell_2} <
150^\circ, \ M_{\ell_1 \ell_2} > 10 \ {\rm GeV \ and} \ \neq M_Z \pm
20 \ {\rm GeV}.
\label{nine}
\ee
Table 2 summarises the signal and background cross-sections after
these cuts, where we have included a $\tau$ identification efficiency of
50\% along with a 0.5\% probability of mistagging a normal hadron jet
as $\tau$ \cite{twelve}.  The latter is a conservative assumption,
since the probability of a normal hadron jet faking a 1-prong
$\tau$-jet with $p_T \sim 20$ GeV has been estimated to be about 0.3\%
for the CDF experiment in Run-1, going up to 0.8\% for the
(1+3)-prongs $\tau$-jet \cite{thirteen}.

\begin{enumerate}
\item[{}] 
Table 2. The signal and background cross-sections (in fb) in
the $\ell \ell \tau$ channel after the cuts of eq. (\ref{nine}).  It
includes a 50\% efficiency factor for $\tau$ identification along with
a 0.5\% probability of mistagging a normal hadron jet as $\tau$.
\end{enumerate}
\[   
\begin{tabular}{|c|c|}
\hline
Signal & Background \\
$\tilde W_1 \tilde Z_2$ & $(Z^\star/\gamma^\star)W$ \ \ \ \ \ \ \ \ \
\ $(Z^\star/\gamma^\star)j$ \\
\hline
& \\
27 & 0.1  \ \ \ \ \ \ \ \ \ \ \ \ \ \ \ \ 0.04 \\
& \\
\hline
\end{tabular}
\]
Thanks to the ${E\!\!\!/}_T$ and the dilepton mass and opening angle
cuts, the potentially large $(Z^\star/\gamma^\star)j$ background is
reduced to $\sim 0.1\%$ of the signal.  We have estimated this
background using a simple analytic formula for the matrix element
neglecting the vector coupling of $Z$ to $\ell\bar\ell$.  The matrix
element for $(Z^\star/\gamma^\star)W$ has been evaluated using
MADGRAPH \cite{fourteen}.

Fig. 1 shows the $P_\tau = +1$ signal as a function of the fractional
$\tau$-jet momentum (R) carried by the charged-prong.  For comparison
it also shows the corresponding distribution assuming the signal to
have $P_\tau = -1$.  This could be the case e.g. in some alternative
SUSY model with a higgsino LSP.  The complimentary shape of the two
distributions, as discussed earlier, is clearly visible in this
figure.  The $P_\tau = +1$ signal shows the peaks at the two ends from
the $\rho_L, a_{1L}$ along with the pion contribution (added to the
last bin), while the $P_\tau = -1$ distribution shows the central peak
due to the $\rho_T, a_{1T}$ along with a reduced pion contribution
\cite{six},\cite{nine}.  The expected luminosity of 2 fb$^{-1}$ per experiment
in Run-2 corresponds to $\sim 54$ signal events in the $\ell^+ \ell^-
\tau$ channel for each experiment without any serious SM background.
Thus one can use this distribution in this case as a confirmatory test
of the minimal SUGRA model.

\begin{figure}[htbp]
\vspace*{-4.9cm}
\hspace*{-3.5cm}
\mbox{\psfig{figure=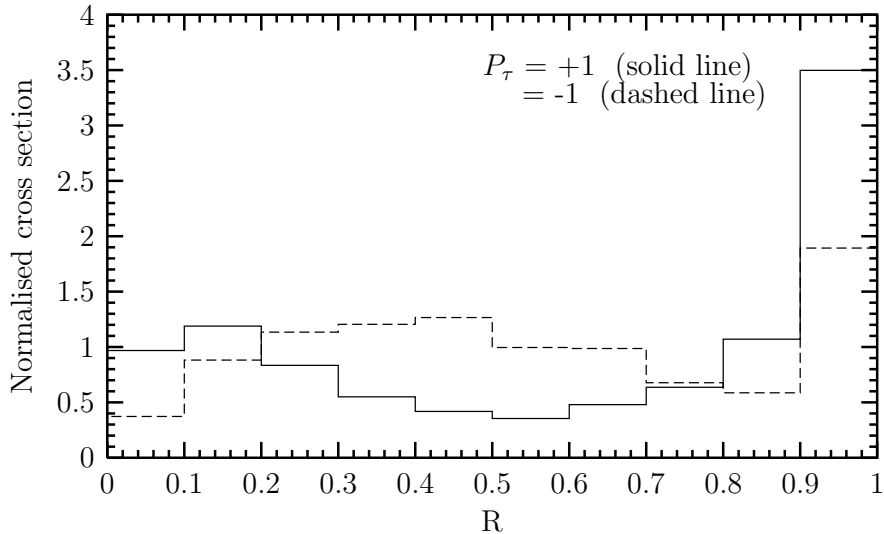,width=20cm}}
\vspace*{-16.7cm}
\caption[Fig.1]{\it
The normalised $\ell\ell\tau$ signal cross-section in the
1-prong hadronic $\tau$-jet channel shown as a function of the
$\tau$-jet momentum fraction (R) carried by the charged prong for
$P_\tau = +1 (-1)$.}
\end{figure}

\subsection*{II.~$m_0 \sim m_{1/2}$ and large $\tan\beta$ ($\tau\tau$
signal):} 
The sparticle spectrum of the bottom row of Table 1 imply
that in this case the dominant decay modes of $\tilde Z_2$ and $\tilde
W_1$ are 
\be
\tilde Z_2 \rightarrow \tau \tilde \tau_1 \rightarrow \tau^+ \tau^-
\tilde Z_1,
\label{ten}
\ee
\be
\tilde W_1 \rightarrow \nu_\tau \tilde\tau_1 \rightarrow \tau \nu_\tau
\tilde Z_1.
\label{eleven}
\ee
Thus one expects a $\tau\tau$ signal from $\tilde W_1 \tilde Z_2$,
$\tilde W_1 \tilde W_1$ and $\tilde \tau \bar{\tilde\tau}$ production
with $P_\tau \simeq 1$ each.  The 1st process contains a 3rd $\tau$ from
$\tilde Z_2 \rightarrow \tau\tilde\tau_1$, whose polarization depends
on the model parameters.  The contribution from the dominant $(\tilde
W)$ component of $\tilde Z_2$ coupling to the subdominant
$(\tilde\tau_L)$ component of $\tilde\tau_1$ has $P_\tau = -1$, while
that from the subdominant $(\tilde B)$ component of $\tilde Z_2$
coupling to the dominant $(\tilde\tau_R)$ component of $\tilde\tau_1$
has $P_\tau = +1$.  And it is the other way around for the higgsino
component of $\tilde Z_2$.  But in any case the $\tau$ resulting from
this decay is relatively soft for the reason mentioned above and
rarely survives the $\tau$-identification cut of $p_T^{\tau-{\rm jet}}
> 15$ GeV.  Therefore we shall require the identification of two
$\tau$ jets with $P_\tau = +1$, while there may be occasionally a 3rd
$\tau$ jet with any polarization (inclusive $\tau\tau$ channel).
We shall neglect the contribution from this 3rd $\tau$ to the signal
cross-section for simplicity, which means a marginal underestimation
of the signal.  The raw cross-sections for $\tilde W_1
\tilde Z_2$, $\tilde W_1 \tilde W_1$ and $\tilde \tau \tilde \tau$
production processes are 770, 850 and 40 fb respectively.  We impose
the following cuts:
\[
p_T^{\tau-{\rm jet}} > 15 \ {\rm GeV}, \ {E\!\!\!/}_T > 20 \ {\rm
GeV}, \ |\eta_{\tau-{\rm jet}}| < 2.5,
\]
\be
30^\circ < \phi_{\tau\tau} < 150^\circ, \ \phi_{{E\!\!\!/}_T \tau} >
20^\circ, \ M_{\tau\tau} > 30 \ {\rm GeV \ and} \ \neq M_Z \pm 30 \
{\rm GeV},
\label{twelve}
\ee 
where we have reconstructed the invariant mass of the $\tau$-pair for
the signal and background events after resolving the ${E\!\!\!/}_T$ into
their respective directions. The reconstructed $M_{\tau\tau}$ represents
the physical invariant mass of the 
$\tau$-pair for the $(Z^\star/\gamma^\star)j$ background; and it plays 
a very effective role in suppressing this 
background. Of course it does not represent the physical $\tau\tau$ 
invariant mass for the signal and other background processes, which 
have additional sources of ${E\!\!\!/}_T$; and the corresponding 
cut does not have any significant effect on these contributions. 
The resulting signal and background cross-sections are listed in Table 3.


We see from the 1st row of Table 3 that the $Wj$
background, with the jet faking as a $\tau$, is about 5 times
larger than the $\tau\tau$ signal.  In view of the importance of this
background we have estimated it via the on-shell $Wj$ as well as the
3-body production processes $q' \bar q (g) \rightarrow \tau\nu g(q)$
using the matrix elements from \cite{fifteen}.  The two estimates agree
to within 5\%.\footnote{As in \cite{four} we have neglected the QCD 
dijet background. They have been estimated to be relatively small after 
imposing the $\tau$-identification and the ${E\!\!\!/}_T$ cuts[3]. As we shall 
see below they can be further suppressed by the R$>$0.8 cut on both the jets.}  

\begin{enumerate}
\item[{}] 
Table 3. The signal and background cross-sections (in fb) in
the $\tau\tau$ channel after the cuts of eq. (\ref{twelve}), including
a 50\% efficiency factor for each $\tau$ along with a 0.5\%
probability for mistagging a normal hadron jet as $\tau$.  The last
row shows the total signal and background cross-sections after the $R > 0.8$
cut on the two $\tau$-jets.
\end{enumerate}
\[
\begin{tabular}{|c|c|}
\hline
Signal & Background \\
$\tilde W_1 \tilde Z_2 \ \ \ \ \ \tilde W_1 \tilde W_1 \ \ \ \ \
\tilde\tau \bar{\tilde\tau}$ & $Wj \ \ \ \ \ (Z^\star/\gamma^\star)j \
\ \ \ \ WW \ \ \ \ \ t\bar t$ \\
\hline
8 \ \ \ \ \ \ \ \ \ \ 8.7 \ \ \ \ \ \ \ 0.5 & 80 \ \ \ \ \ \ \ \ \ \ \
2 \ \ \ \ \ \ \ \ \ \ \ 1.7 \ \ \ 1.2 \\
\hline
3.5 & 3.6 \\
\hline
\end{tabular}
\]

Fig. 2 compares the $P_\tau = +1$ signal and this $P_\tau = -1$
background as functions of the $\tau$-jet momentum fraction $R$
carried by the charged prong.  It clearly shows the complimentary
shapes of the two distributions, similar to those of Fig. 1.  It means
that the difference comes mainly from the opposite polarizations of
$\tau$ rather than kinematic difference between the signal and the
background.  Requiring the charged track to carry $> 80\%$ of the
$\tau$-jet energy-momentum $(R > 0.8)$ retains 45\% of the signal as
against only 20\% of the background.  Moreover the $R > 0.8$ cut is
also known to reduce the fake background from normal hadron jets by at
least a factor of 5 \cite{sixteen}.  Thus demanding both the
$\tau$-jets to contain hard charged tracks, carrying $> 80\%$ of their
momenta, would reduce the signal by a factor of 5 while reducing the
dominant background by at least a factor of 25. The same is true for
the QCD dijet background not considered here. The $\tau\tau$
background from $WW$ and $t\bar t$ are also reduced by a factor of 25
each. On the other hand the background from $Z^\star/\gamma^\star
\rightarrow \tau\tau$ has $P_\tau = 0$, and the corresponding
distribution lies midway between those of $P_\tau = \pm 1$.  The
resulting suppression factor is $\simeq (1/3)^2$.
\begin{figure}[htbp]
\vspace*{-5.5cm}
\hspace*{-3.5cm}
\mbox{\psfig{figure=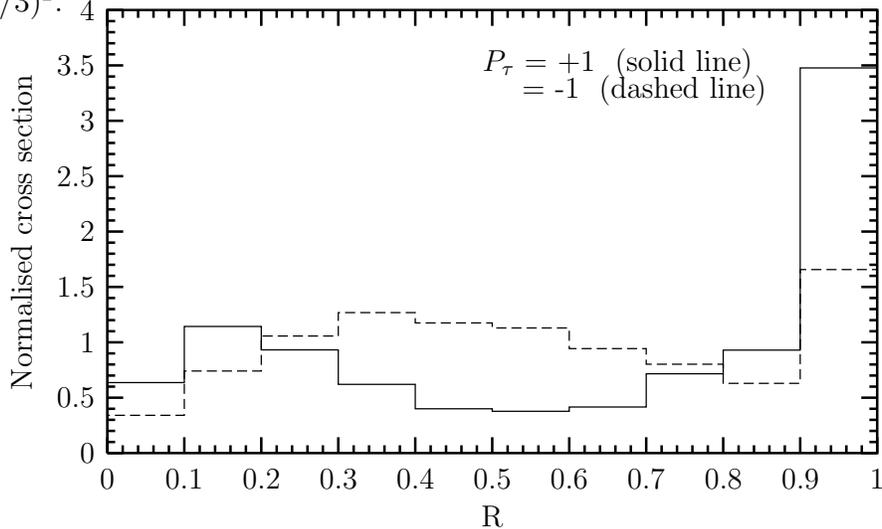,width=20cm}}
\vspace*{-16.7cm}
\caption[]{\it
The normalised SUSY signal $(P_\tau = +1)$ and $Wj$
background $(P_\tau = -1)$ cross-sections in the 1-prong hadronic
$\tau$-jet channel shown as functions of the $\tau$-jet momentum
fraction (R) carried by the charged prong.}
\end{figure}

As we see from the bottom row of Table 3 the $R > 0.8$ cut on both the
$\tau$-jets reduces the total background to the signal level,
i.e. about 3.5 fb each.  With the expected Run-2 luminosity of
2 fb$^{-1}$ per experiment, one expects a combined yield of $\sim 14$
signal events against a similar background from CDF and D${\rm
O\!\!\!\!/}$.  Note that the corresponding significance level is
$S/\sqrt{B} \simeq 4$ with or without the $R > 0.8$ cut.  Nonetheless
it is no mean gain that this cut can enhance the signal to background
ratio from 1/5 to at least 1.  This means that the $\tau\tau$ channel
can offer a viable SUGRA signature along with the $\ell^+\ell^-\tau$
channel at the Tevatron upgrades, starting with Run-2.  It may be
noted from Table 3 that requiring the $\tau$ pair to have opposite
sign (same sign) will retain a little over $3/4$ (under 1/4) of
the signal while retaining 1/2 of the dominant background.  Thus with
sufficient luminosity it may be possible to improve the signal to
background ratio by requiring the $\tau$ pair to have opposite sign.
Finally it should be noted that while we have focussed the current
analysis on the SUGRA model the same polarization strategy can be used
to distinguish the SUSY signal from the SM background in the gauge
mediated SUSY breaking model \cite{seventeen}, where one expects a
$P_\tau = +1$ from the $\tilde\tau_R \rightarrow \tau \tilde G$ decay.

We are grateful to Dhiman Chakraborty, Daniel Denegri and especially
Manuel Drees for many helpful discussions.  DPR acknowledges partial
financial support from IFCPAR.


\bigskip

\begin{thebibliography}{999}
\bibitem{one} Review of Particle Properties, Euro. Phys. J. C15, 1
(2000). 
\bibitem{two} H. Baer, M. Drees, F. Paige, P. Quintana and X. Tata,
Phys. Rev. D61, 095007 (2001).
\bibitem{three} H. Baer, C. Chen, M. Drees, F. Paige and X. Tata,
Phys. Rev. Lett. 79, 986 (1997); Phys. Rev. D58, 075008 (1998);
Phys. Rev. D59, 055014 (1999).
\bibitem{four} K.T. Matchev and D.M. Pierce, Phys. Rev. D60, 075004
(1999); J.D. Lykken and K.T. Matchev, Phys. Rev. D61, 015001 (2000).
\bibitem{five} V. Barger, C. Kao and T. Li, Phys. Lett. B433, 328
(1998); V. Barger and C. Kao, Phys. Rev. D60, 115015 (1999). See also
V. Barger et al., hep-ph/0003154.
\bibitem{six} S. Raychaudhuri and D.P. Roy, Phys. Rev. D52, 1556
(1995); D53, 4902 (1996).
\bibitem{seven} D.P. Roy, Phys. Lett. B459, 607 (1999).
\bibitem{eight} J.H. Kuhn and A. Santamaria, Z. Phys. C48, 445
(1990). 
\bibitem{nine} B.K. Bullock, K. Hagiwara and A.D. Martin,
Phys. Rev. Lett. 67, 3055 (1991); Nucl. Phys. B395, 499 (1993). 
\bibitem{ten} H. Baer, F. Paige, S.D. Protopopescu and X. Tata,
hep-ph/0001086. 
\bibitem{eleven} M. Nojiri, Phys. Rev. D51, 6281 (1995).
\bibitem{twelve} CDF Collaboration: F. Abe et. al.,
Phys. Rev. Lett. 79, 3585 (1997); M. Hohlmann, preprint
Fermilab-conf-98-376-E. 
\bibitem{thirteen} M. Hohlmann, Univ. of Chicago, Ph.D. Thesis
(1997). 
\bibitem{fourteen} MADGRAPH, T. Stelzer and W.F. Long,
Comp. Phys. Commun. 81, 357 (1994); HELAS by H. Murayama, I. Watanabe
and K. Hagiwara, KEK-91-11 (1992).
\bibitem{fifteen} V. Barger and R.J.N. Phillips, Collider Physics,
Addison-Wesley (1987).
\bibitem{sixteen} R. Kinnunen and D. Denegri, CMS Note 1999/037; and
private communication from D. Chakraborty and D. Denegri.
\bibitem{seventeen} D.A. Dicus, B. Dutta and S. Nandi, Phys. Rev. D56,
5748 (1997); B. Dutta, D.J. Muller and S. Nandi, Nucl. Phys. B544, 451
(1999). 
\end{thebibliography}
\end{document}